# Qubit guidelines for solid-state spin defects

Gary Wolfowicz[1,2*†], F. Joseph Heremans[1,2*], Christopher P. Anderson[2,3*], Shun Kanai[4-7], Hosung Seo[8], Adam Gali[9,10], Giulia Galli[1,2,11], David D. Awschalom[1,2,3]

[1]*Center for Molecular Engineering and Materials Science Division, Argonne National Laboratory, Lemont, IL 60439, USA*

[2]*Pritzker School of Molecular Engineering, University of Chicago, Chicago, IL 60637, USA*

[3]*Department of Physics, University of Chicago, Chicago, IL 60637, USA*

[4]*Laboratory for Nanoelectronics and Spintronics, Research Institute of Electrical Communication, Tohoku University, Sendai 980-8577, Japan*

[5]*Division for the Establishment of Frontier Sciences, Tohoku University, Sendai 980-8577, Japan*

[6]*Center for Science and Innovation in Spintronics, Tohoku University, Sendai 980-8577, Japan*

[7]*Center for Spintronics Research Network, Tohoku University, 2-1-1 Katahira, Aoba-ku, Sendai 980-8577 Japan*

[8]*Department of Physics and Department of Energy Systems Research, Ajou University, Suwon, Gyeonggi 16499, Republic of Korea*

[9]*Wigner Research Centre for Physics, PO Box 49, H-1525, Budapest, Hungary*

[10]*Department of Atomic Physics, Budapest University of Technology and Economics, Budafoki ut 8., H-1111 Budapest, Hungary*

[11]*Department of Chemistry, University of Chicago, Chicago, IL 60637, USA*

[†]*email: gwolfowicz@anl.gov*

*These authors contributed equally to this work

## Abstract

Defects with associated electron and nuclear spins in solid-state materials have a long history relevant to quantum information science going back to the first spin echo experiments with silicon dopants in the 1950s. Since the turn of the century, the field has rapidly spread to a vast array of defects and host crystals applicable to quantum communication, sensing, and computing. From simple spin resonance to long-distance remote entanglement, the complexity of working with spin defects is fast advancing, and requires an in-depth understanding of their spin, optical, charge, and material properties in this modern context. This is especially critical for discovering new relevant systems dedicated to specific quantum applications. In this review, we therefore expand upon all the key components with an emphasis on the properties of defects and the host material, on engineering opportunities and other pathways for improvement. Finally, this review aims to be as defect and material agnostic as possible, with some emphasis on optical emitters, providing a broad guideline for the field of solid-state spin defects for quantum information.

## Introduction

Defects in the solid-state are at the center of countless challenges and opportunities in condensed matter physics. They are either highly detrimental, e.g. to crystalline growth, or beneficial due their ability to modulate and control material properties. When isolated, impurities stand as analogues of atomic systems in an effective "semiconductor vacuum"[1] with properties defined by the host substrate, leading to their study for semi-classical and later quantum applications.

In 1958, electron spin resonance experiments were realized with phosphorus dopants in both natural and isotopically purified silicon, with the first spin-echo experiments showing coherence times as high as 0.5 ms[2]. In the 1960s, the first solid-state laser used chromium dopants in ruby[3], while many electron spin resonance experiments were realized in rare-earth ions in oxides[4]. It was not until the turn of the century however, that a strong push for quantum applications started. This began with the proposal of quantum computers based on donors in silicon with electrical gates[5] or based on rare-earth ions in $Y_2SiO_5$ with optical cavities[6], and with the first measurement of single nitrogen-vacancy (NV) defects in diamond[7]. Following these successes, defects with spins in the solid-state have rapidly been applied to all three major fields of quantum science: sensing, computing, and communication. More recently, novel defects such as group IV dopant-vacancies in diamond[8] or vacancy complexes in silicon carbide[9] have emerged and are especially promising for quantum communication.

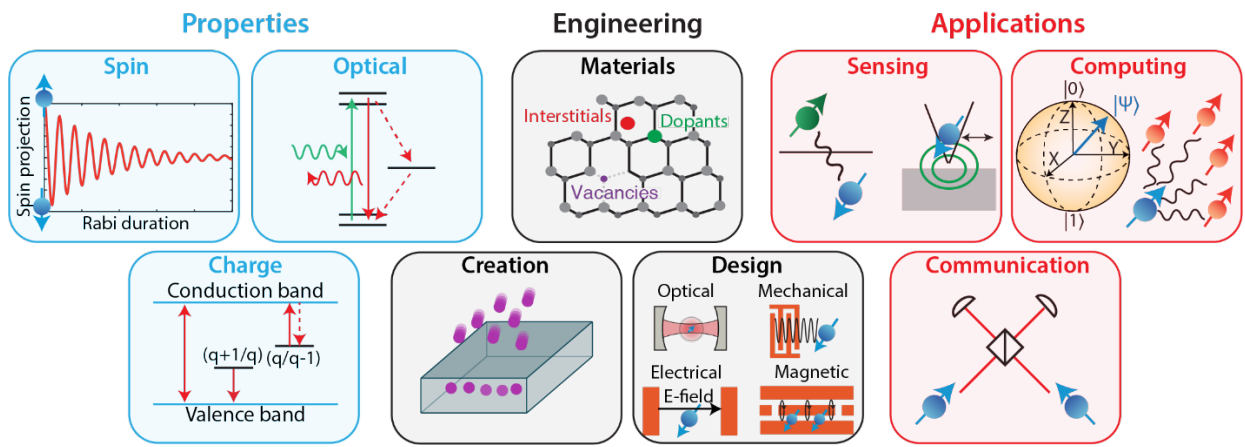

Figure 1 | **Spin defects in the solid-state for quantum information science.** Illustration of key concepts including the three major defect properties (spin, optical, and charge) in the blue panels, engineering considerations (materials, creation, and design) in the black panels, and the three major quantum applications (sensing, communication, and computing) in the red panels. Charge: transition energy levels for charge conversion. Spin: spin projection during a Rabi oscillation. Optical: pumping and photoluminescence in a defect with an inter-system crossing. Material: various defect types and lattice sites. Creation: defect creation by implantation or irradiation. Design: optical, mechanical, electrical, and magnetic devices for interfacing with spin defects. Sensing: electron spin as sensor for other local spins or magnetic fields. Communication: entanglement between two spins via two-photon interference. Computing: interaction of an electron spin with many nuclear spin registers.

Defects are defined by their spin, optical, and charge states as well as by the properties of their host material (Fig. 1). It is the interplay between all these components that allows complex experiments and technologies to be realized. Quantum sensing, for example, uses the spin state to acquire a phase shift from interactions with the environment[10], while the optical interface (i.e. spin-to-photon conversion) allows optical readout of the spin qubit, potentially enhanced by spin-to-charge quantum non-demolition (QND) measurements[11]. Quantum computing is most closely realized with large clusters of nuclear spin registers coupled to an electron spin[12,13], alongside optical or charge-based control for efficient initialization and readout[14]. Quantum communications demands an efficient spin-dependent optical interface[15] and a spin quantum memory[16].

In this guide, we broadly consider electron and nuclear spins associated with point defects primarily in bulk solid-state materials in the context of quantum science, with applicability to recent developments including spins in 2D materials[17,18]. This encompasses single-atomic and atomic-vacancy defects as well as vacancy complexes with magnetic, electrical, mechanical, and optical interfaces for initialization, readout, and control. The manuscript is divided into four main sections: I) Spin properties, II) Optical properties, III) Charge properties, and IV) Material considerations. For completeness, we refer to other more focused reviews based on defects in diamond[8,19,20], silicon carbide[9,21], silicon[22], rare-earth dopants[23], theoretical modeling[24] and other aspects[25–27].

## Spin properties

For defects in the solid-state, quantum information is generally encoded in the electron spin of the orbital ground state of the defect (with some exceptions[28]). Electron spins provide a controllable qubit with long relaxation and coherence times. They can be coupled to nuclear spins for very long quantum memories and advanced (e.g. QND) applications. The spin state is therefore central to the quantum hardware and has been the subject of most theoretical and experimental research.

*Spin relaxation*

The spin relaxation time, $T_1$, is the characteristic time for the spin to reach an equilibrium state after random spin flips along the spin quantization axis. $T_1$ fundamentally limits the possible coherence times such that $T_2 \leq 2T_1$, though in practice $T_2 \leq 0.5T_1$ or $1T_1$[29,30]. The electron spin $T_1$ in the solid-state, our focus here, is predominantly set by spin-lattice relaxation, namely thermal relaxation from absorption, emission or scattering of phonons in the crystal through the spin-orbit interaction. This sets the operating temperature regime of the qubit, as shown in Fig. 2a for common materials.

Predicting $T_1$ is challenging and requires *ab initio* computations of the spin-flip matrix elements from the spin-phonon interaction potential[31,32]. However, its dependence on various parameters such as the lattice temperature $T$ has been well described[33,34] and experimentally verified in many systems[35–37]. $T_1$ models are generally divided into three relaxation mechanisms (Fig. 2b): direct absorption or emission of one phonon resonant between two electron spin states, Raman processes by virtual absorption and emission of two phonons, and Orbach[38] relaxation by phonon excitation to a higher excited state followed by decay and emission of a phonon.

For non-integer (Kramers) spins, $T_1$ follows[33,34]:

$$\frac{1}{T_1} \approx K_D K_\theta \Delta E_{\text{spin}}^4 T + K_\theta^2 (K_{R1} T^5 + K_{R2} \Delta E_{\text{spin}}^2 T^7 + K_{R3} T^9) + K_O K_\theta \frac{\Delta E_{\text{orb}}^3}{\exp\left(\frac{\Delta E_{\text{orb}}}{k_b T}\right) - 1}, \quad (1)$$

and for integer (non-Kramers) spins, $T_1$ follows:

$$\frac{1}{T_1} \approx K_D K_\theta \Delta E_{\text{spin}}^2 T + K_\theta^2 (K_{R1} T^5 + K_{R2} T^7) + K_O K_\theta \frac{\Delta E_{\text{orb}}^3}{\exp\left(\frac{\Delta E_{\text{orb}}}{k_b T}\right) - 1} \quad (2)$$

where $K_\theta \propto \rho^{\frac{2}{3}}/\theta_D^5$ is a host dependent parameter with $\rho$ the atomic density and $\theta_D$ the Debye temperature. $K_D$, $K_{R1-3}$ and $K_O$ are respectively coefficients for the direct, Raman and Orbach mechanisms and are related to the spin-phonon coupling, predictable from first principles[39,40]. $\Delta E_{\text{spin}}$ is the energy splitting between the ground spin states, $\Delta E_{orb}$ is the energy splitting (lower than the Debye frequency) between the ground and a nearby orbital excited state, and $k_b$ is the Boltzmann constant. Eq. 1 and 2 are valid for $\Delta E_{spin}/k_B < T < \theta_D$.

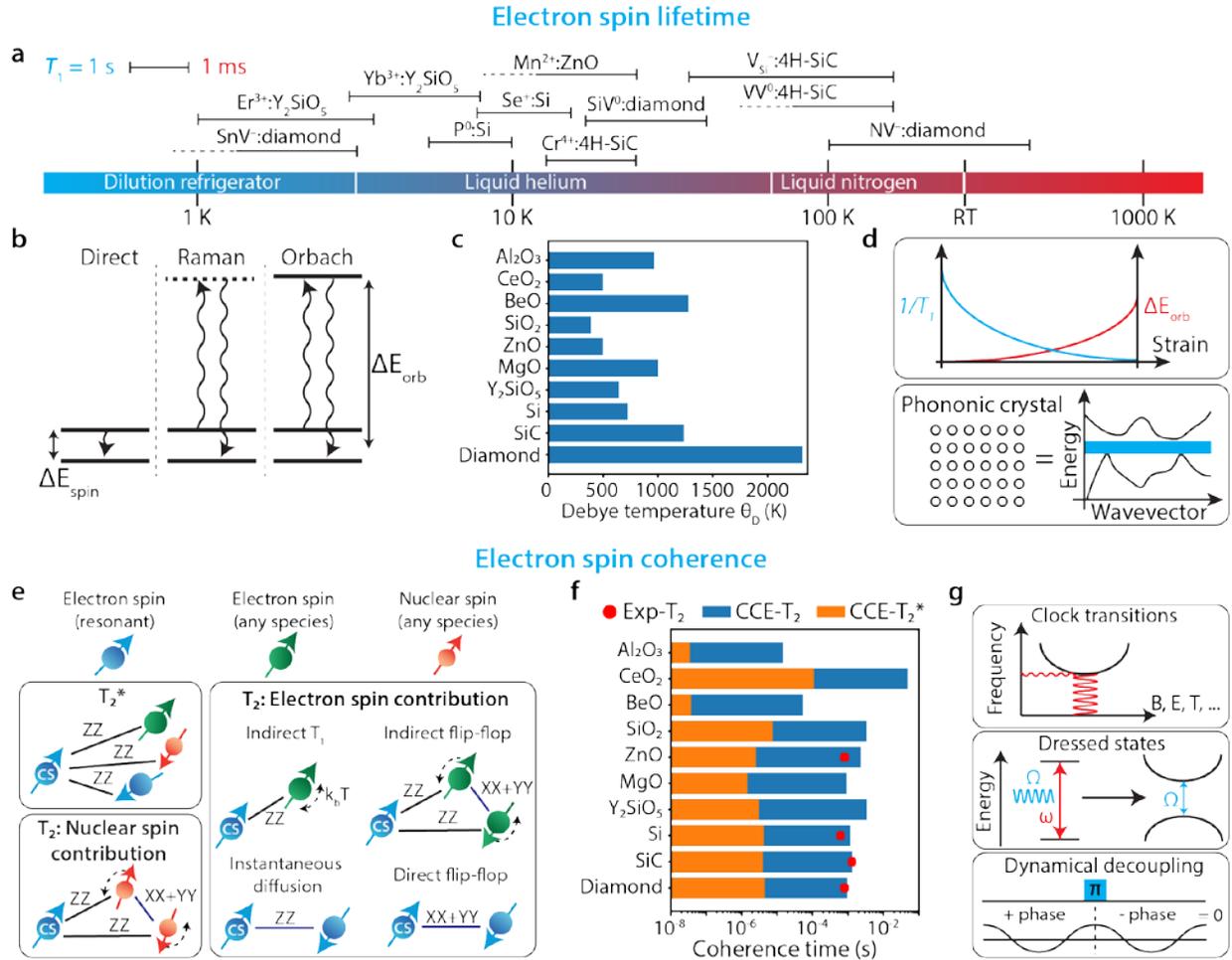

Figure 2 | **Electron spin relaxation and coherence. a|** Temperature ranges corresponding to $T_1$ times from 1 ms to 1 s in common defects for quantum information[36,41–50]. Dotted lines are for missing parameters from the literature. **b|** Phonon processes for spin-lattice relaxation of electron spins in the solid-state. $\Delta E_{spin}$ and $\Delta E_{orb}$ are described in the main text. **c|** Summary of Debye temperatures in relevant qubit host materials[51–55]. **d|** Engineering $T_1$ relaxation. Top: strain tuning of the ground state orbital splitting to reduce the Orbach process. Bottom: phononic structure to gap (in blue) relevant energies such as $\Delta E_{orb}$ or $\Delta E_{spin}$. **e|** Dominant decoherence mechanisms of an electron spin in the solid-state. XX, YY and ZZ represent the interaction axis, with Z the quantization axis. **f|** CCE calculations of $T_2^*$ and $T_2$ (in bar) for a spin-1/2 electron spin coupled to a nuclear spin bath in various host materials with natural isotope abundance. The only defect-specific parameter used is the standard gyromagnetic ratio $\gamma = 28$ GHz/T. The hyperfine contact term is neglected, resulting in discrepancies for defects in silicon, for example. The red dots are experimental values for Mn$^{2+}$:ZnO[41], P$^0$:Si[56], VV$^-$:4H-SiC[57] and NV$^-$:Diamond[30] which includes $S = 1$ systems (expected ~20% variations from $S = 1/2$ calculations). **g|** Hamiltonian engineering of the electron spin to create transitions insensitive to external fluctuations such as magnetic field (B), electric field (E), or temperature (T). From top to bottom: Clock transitions by static tuning of the spin interactions, dressed states for dynamical creation of clock transitions, and dynamical decoupling for cancelling undesired phase accumulations.

For material design (see Fig. 2c), the choice of the Debye temperature $\theta_D$ (via $K_\theta$) has a large impact on the spin relaxation, especially for Raman processes ($\propto K_\theta^2$). For the latter, one power law generally dominates such as $T^9$ in rare-earths ions[58], while the Orbach mechanism often dominates in the presence of a low lying orbital state[37,58–60]. At extremely low temperatures (<< 1 K), a phonon bottleneck effect can occur where there are not enough phonons to transfer energy from the spin to the lattice and the relaxation time is reduced[34,61]. Finally, the dependence on $\Delta E_{\text{spin}}$ for some of the mechanisms allows for tunability, usually using an external magnetic field[35].

The electron spin relaxation is not limited to phonon processes. Resonant magnetic or electric noise from the environment can effectively drive random flips of the defect electron spin[62,63]. In particular, this occurs near surfaces from dangling bonds[64,65] (see section Spin coherence), and in turn, monitoring $T_1$ can be used directly as a sensing mechanism[66,67]. Another decay pathway is via charge instability where the defect alternates between charge states in a non-spin-conserving manner[62]. Finally, undesired photoexcitation to an excited state followed by relaxation may cause spin flips similar to phonon-induced processes.

An isolated nuclear spin-1/2 in the solid-state has very few direct phonon-mediated $T_1$ mechanisms. Its spin relaxation time is generally limited by coupling with the rest of the nuclear spin bath or with the electron spins in a sample. For the latter case, the nuclear spin $T_1$ is limited by cross-relaxation involving simultaneous electron and nuclear spin flips driven by phonon modulation of the hyperfine coupling[36]. In this case, the nuclear spin coherence may have an upper limit $T_{2,\text{nuclear}} \leq 2T_{1,\text{electron}}$[16,68], though this can be mitigated via dissipative decoupling[69].

The phonon-limited $T_1$ is one of the hardest characteristics to improve for a given temperature, defect, and host material. However, the Orbach contribution can be reduced by increasing $\Delta E_{\text{orb}}$ through strain tuning[59], and engineering a phononic bandgap around $\Delta E_{\text{spin}}$ or $\Delta E_{\text{orb}}$ could eliminate either the direct or Orbach processes (Fig. 2d). Nanostructures smaller than the wavelength of relevant acoustic phonons could theoretically improve $T_1$ for spins, though they may suffer from surface proximity effects. Finally, lowering $T_1$ may be desired for spin polarization by thermal relaxation, and can be realized by Purcell enhancement of the spontaneous emission in a microwave cavity[70], normally negligible at microwave frequencies ($T_1^{spont} \approx 10^{17}$ s at 1 GHz[71]).

The substantial $T_1$ times of solid-state defects are one of the core advantages over other quantum systems. We summarize the following key points:

- The electron spin $T_1$ fundamentally limits the maximum electron spin coherence time.
- A high Debye temperature is preferable for qubit operation at elevated temperatures.
- Resonant noise sources can also drive spin flips.
- Improving $T_1$ remains challenging but may be achieved for systems limited by direct or Orbach relaxation.

*Spin coherence*

Decoherence is the loss of phase information in a quantum system. Quantum applications require long coherence times for longer memories, higher control fidelities, and longer phase acquisition times. Thankfully, decoherence can be well predicted and suppressed with sufficient knowledge of the quantum system and its environment. Typically, spin qubits lose their phase coherence from surrounding fluctuating magnetic sources (i.e. other nuclear and electron spins). In prevalent materials, the electron spin coherence time naturally starts from milliseconds and can be extended by several orders of magnitude to seconds[30,72–74].

Coherence is characterized by two key figures: the inhomogeneous dephasing time $T_2^*$ and the homogeneous (or Hahn echo) dephasing time $T_2$. In spin ensembles, $T_2^*$ originates mainly from the random and static distribution of spin states and their interaction with the environment (bath). For single spins, $T_2^*$ results mainly from experiment-to-experiment fluctuations in the spin bath state during averaging. In singles and ensembles, $T_2$ is measured using a refocusing $\pi$ pulse to suppress the static and slow fluctuations, and it is therefore set by fast noise processes.

In Fig. 2e, we consider the most common decoherence mechanisms for an electron spin in the solid-state. The naturally abundant non-zero nuclear spin isotopes of a material are the dominant sources of magnetic field noise and can be readily simulated using fully quantum mechanical cluster correlation expansion (CCE) calculations[75,76]. From the spin Hamiltonian, one computes the coherence function $L(t) = \text{Tr}[\rho(t)S^+]/\text{Tr}[\rho(0)S^+]$ for an electron spin coupled to nuclear spins randomly distributed in the lattice, where $t$ is the time, $\rho$ is the density operator and $S^+$ is the electron spin raising operator. $T_2$ is then obtained from the decay profile $L(t) = e^{-(t/T_2)^n}$ after ensemble averaging, with $n$ the stretching exponent[77]. The density operator is approximated considering different orders of nuclear spin cluster sizes[57,78]. In Fig. 2f, we present CCE calculations[79] of predicted $T_2^*$ and $T_2$ for relevant qubit host materials. There are three material factors that can reduce this decoherence contribution: a low abundance of non-zero nuclear spin isotopes, low nuclear gyromagnetic ratios, and host lattices that prevent nearest-neighbor coupling of nuclear spins of the same species.

Semi-classical methods can also be used to predict coherence based on the random distribution (valid at low spin bath concentration $C_B$) of spin states in the electron or nuclear spin bath. A coarse upper bound estimate for $T_2^*$ is $1/R_{\text{dipolar}}$, with $R_{\text{dipolar}}$ the characteristic dipolar coupling rate (in units of frequency)[71,80,81]:

$$R_{\text{dipolar}} = C_B(2\pi\gamma)(2\pi\gamma_B)\frac{\pi}{9\sqrt{3}}\mu_0\hbar \qquad (3)$$

with $\gamma$ and $\gamma_B$ the measured and bath spin gyromagnetic ratios, respectively.

Surrounding electron spins start to significantly (< seconds) contribute to $T_2$ when their concentration reaches about $10^{12} - 10^{14}$ cm$^{-3}$. First, "instantaneous diffusion" occurs when the measured electron spin and other (usually electron) bath spins are simultaneously on resonance with a $\pi$ pulse. In this case, the pulse has no refocusing effect and decoherence ensues like $T_2^*$, with $T_2 = 1/R_{\text{dipolar}}$ and usually $\gamma = \gamma_B$. Instantaneous diffusion rarely dominates in samples with isolated single spins due to the low concentration of resonant spins.

Magnetic fluctuations can also arise from the lifetime ($T_{1,B}$) of electron spins in the bath. This "indirect $T_1$" contribution to the decoherence follows the relation[72,82] $1/T_2 = \sqrt{R_{\text{dipolar}}/2\pi T_{1,B}}$. Finally, the spin state exchange, or flip-flop, between resonant spins becomes relevant at higher spin concentration ($1/T_2 \propto C_B^2$)[83], though can be suppressed by local detuning such as magnetic field gradients. Flip-flops are either direct when involving the measured spin, or indirect when only within bath spins. Direct flip-flop may contribute as a $T_1$ mechanism: if the state exchange includes a spin outside of the measured spin ensemble, or occurs for single spin measurements, then the spin projection information is lost after the spin-flip[84].

Electric field noise can also be a major source of decoherence via modulation of certain spin parameters (see section Spin control)[85–87]. Large electrical noise is rarely present in bulk materials and appears instead in nanoscale devices with metallic or highly doped regions, or near surfaces with dangling bonds[88]. $T_2^*$ is also susceptible to static variations from strain in the crystal. Finally, decoherence can also occur through rapid thermal excitation/relaxation into an excited state with spin parameters different from those of the ground state[89].

An increase of coherence time can be engineered by reducing the nuclear spin bath through isotopic purification[73,90–92] and by reducing the concentration of electron spins during crystal growth, defect creation, and doping. Therefore, finding host materials where this is possible is a key criterion for quantum information. Tuning the dimensionality (2D vs bulk) of the material can also improve decoherence from the different spatial distribution of spins[93]. Similarly, for a dipole-coupled spin bath, the magnetic fluctuations can be suppressed for specific magnetic field orientations depending on the host lattice structure[56].

In parallel, a defect's spin can be tuned to be less sensitive to its local environment by finding "clock" (or zero first-order Zeeman[94]) transitions, or by dynamically creating a decoupled spin subspace, as illustrated in Fig. 2g. Clock transitions arise in systems with at least two competing spin interactions, e.g. Zeeman and hyperfine[73,95] or zero-field interactions[74], such that transition frequencies $f$ become locally independent of a tunable parameter $P$, i.e. $df/dP \to 0$ (to first order). For a magnetic field $P = B$, this gives an effective gyromagnetic ratio $\gamma_{\text{eff}} = df/dB$ that reduces the interaction with the spin bath (and Eq. 3 for example)[96].

While clock transitions passively decouple a qubit from its environment, the same decoupling effect can be obtained dynamically. Under a continuous resonant drive, the spin states are dressed and may form an avoided crossing split by the Rabi amplitude, and with a local frequency minimum similar to a clock transition[97,98]. The drive can also be pulsed in a dynamical decoupling sequence. Dynamical decoupling is a generalization of the Hahn echo to arbitrary pulse sequences that cancel specific interactions with the environment[30,99], and is usually understood as a classical noise filter[100]. Unfortunately, decoupling techniques are limited when correcting for strong or resonant noise, where homogeneous and inhomogeneous errors in drive frequency or amplitude are present, and may cause drive-induced heating[101].

In summary, key considerations are:

- Magnetic field noise from the nuclear spin bath which can be mitigated for host materials with low spin-full isotope concentrations.
- Predicting coherence using CCE and understanding through semi-classical models.
- Multiple competing spin interactions which provide clock transitions.
- The improvement of coherence using dressed-states and dynamical decoupling for systems with low spin control errors.

*Spin control*

A large variety of protocols are available to coherently control the electron (or nuclear) spin state[25]. These schemes include optical, magnetic, electric, and strain fields, each with advantages and disadvantages depending on the specific electron spin and orbital level structures of the defect, and the application at hand.

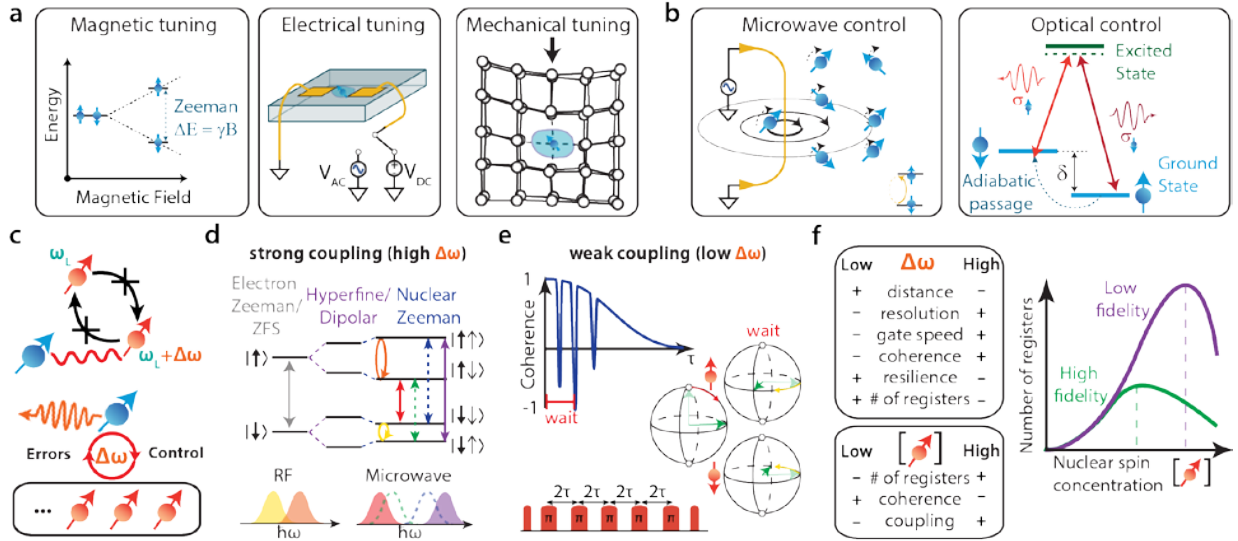

Figure 3 | **Electron and nuclear spin control. a|** Electron spin frequency tuning via external magnetic, electrical, and mechanical fields. **b|** Electron spin driving using magnetic (or other) fields at microwave frequencies (left). Optical control via an orbital excited state (right). **c|** Electron spin coupling to nearby nuclear spin registers in the environment. The detuning $\Delta\omega$ of the coupled nuclear spin from the rest of the nuclear spin bath is a key parameter that reduces interactions with the bath, that also sets control mechanisms and accumulated errors on the nuclear register. **d|** Nuclear spin driving schemes using resolved conditional rotations for strongly coupled spins (high $\Delta\omega$). The dotted lines are forbidden transitions. **e|** Ramsey and dynamical decoupling schemes for driving weakly coupled nuclear spins (low $\Delta\omega$) where nuclear spin-dependent phases are accumulated and projected. **f|** Trade-offs involved in the choice of hyperfine strength ($\Delta\omega$) and nuclear spin concentration (left). The nuclear spin bath concentration must be chosen carefully for high fidelity control and the number of nuclear spin registers (right)[91].

An electron spin ($S$) coupled to a nuclear spin ($I$) can be described by the following spin Hamiltonian:

$$H = \underbrace{\mu_B \mathbf{B}\cdot \mathbf{g}_e \cdot \mathbf{S}}_{electron\,Zeeman} + \underbrace{\mathbf{S}\cdot \mathbf{D} \cdot \mathbf{S}}_{zero-field} - \underbrace{\mu_N \mathbf{B}\cdot \mathbf{g}_n \cdot \mathbf{I}}_{nuclear\,Zeeman} + \underbrace{\mathbf{S}\cdot \mathbf{A} \cdot \mathbf{I}}_{hyperfine} + \underbrace{\mathbf{I}\cdot \mathbf{Q} \cdot \mathbf{I}}_{quadrupole} \quad (4)$$

where each bracket denotes a different spin interaction parametrized by the external magnetic field ($\mathbf{B}$), the electron and nuclear spin g-tensor ($\mathbf{g}_e, \mathbf{g}_n$), the zero-field splitting tensor ($\mathbf{D}$) for $S > 1/2$, the hyperfine coupling tensor ($\mathbf{A}$, including both contact and dipolar terms), and the quadrupole tensor ($\mathbf{Q}$) for $I > 1/2$[71]. $\mu_B$ and $\mu_N$ are the Bohr and nuclear magneton, respectively. The microscopic origin of each parameter is well known and their tensorial form can be predicted by *ab initio* calculations[93,102–107].

In addition to the magnetic field $\mathbf{B}$ in Eq. 4, all the interaction tensors and transition frequencies can be controlled by local electric fields ($\mathbf{E}$) and strain ($\boldsymbol{\varepsilon}$) which perturb the electronic wavefunction[108], as illustrated in Fig. 3a. However, the interaction tensors are affected by $\mathbf{E}$ and $\boldsymbol{\varepsilon}$ in different ways, depending on the symmetry and spin of the defect[109–111]. Electron spin resonance typically occurs at microwave (GHz) frequencies largely defined by the electron Zeeman interaction and the zero-field splitting. Practically, the spin's transition frequency should be large compared to the Rabi frequency, to prevent limitations from the rotating wave approximation[112], yet low enough to avoid microwave losses and instrumentation difficulties. Magnetic fields provide the simplest coherent spin control of $\Delta m_s = \pm 1$ transitions using microwave striplines and resonators (see Fig. 3b), though it is challenging to independently manipulate nearby spins. This may be solved using magnetic field gradients or other local detuning methods[5].

Alternatively, nearby spins can be selectively addressed and driven with electric fields confined in nanodevices. Electric field modulation of the zero-field tensor results in $\Delta m_s = \pm 2$ transitions[113], while electric field modulation of the hyperfine interaction results in $\Delta m_s + \Delta m_I = 0$ (flip-flop) transitions[85]. Recent demonstrations have also shown electric driving of the quadrupolar interaction for nuclear spins in Sb dopants in silicon[114].

Mechanical (phonon) control of the spin state works by creating local crystallographic strains (see Fig. 3a), e.g. in a cantilever or acoustic resonator, which allow for full ground state control of both the $\Delta m_s = \pm 1$ and $\Delta m_s = \pm 2$ transitions[109]. Typically, the spin-strain coupling is small. Nonetheless, a strong coupling is desired for quantum transduction between spins and phonons[115], at the potential trade-off of reduced $T_1$.

For a given two-level system, the longitudinal component of the interaction dipole provides tuning of the spin frequency, though at the expense of sensitivity to noise. Coherent qubit control is commonly achieved by resonantly driving the transverse components of the interaction. For example, this makes defects with anisotropic g-factors such that $g_{zz} \ll g_{xx,yy}$ interesting for balancing fast driving speeds with low decoherence, but reduced frequency tuning. Ideally, a spin should be insensitive to fields that are prominent and uncontrolled (noise), but sensitive to fields that are generated for control or for sensing. In addition, achieving high-fidelity control can prove challenging due to decoherence processes and inhomogeneous effects induced by the local environment. Techniques such as optimal control[116], adiabatic passage[117], or composite pulses[118] aid in mitigating these effects.

Ground state electron spin manipulation is also available by optical excitation, which utilizes the orbital excited state levels as an intermediary, e.g. in Λ-like and V-like systems (see Fig. 3b). This enables the use of optical techniques developed in the context of trapped ions and cold atoms[119], including coherent population trapping (CPT) and electromagnetically induced transparency (EIT)[120,121]. These methods are based on coherent dark dressed states, allowing for optical initialization, manipulation, and readout[122–124]. Furthermore, two-qubit gates between two defect's electron spins can be mediated by light in a photonic cavity[125].

For optical control, excited state effects (such as the optical lifetime, coherence, and spectral hopping) are all major sources of errors that limit control fidelities[126]. Many methods have been developed to mitigate these issues including stimulated Raman adiabatic passage (STIRAP), geometric phase and holonomic control[122,126,127], and superadiabatic approaches[128]. Optical control offers localized spin driving limited by the optical spot-size of the excitation laser, or with sub-diffraction control when combined with spectral resolution[129,130].

In summary, understanding the spin Hamiltonian and defect-host coupling is critical toward achieving high-fidelity coherent control. A few key features include:

- The tuning of spin Hamiltonian components through magnetic fields, electric fields, and strain.
- The need to balance coherent control and decoherence with noise.
- Specific defect (e.g. symmetry, g-factor, ZFS) and host properties (e.g. acoustoelastic, piezoelectric) for particular applications.
- Excited orbitals as alternative pathways for ground state spin control.

*Nuclear spin registers*

Besides being a source of decoherence for the defect's electronic spin, nuclear spins can act as key components for quantum communications[27,131,132], computation[133,134], and sensing[135]. Due to the low magnetic moment of nuclear spins and weak interactions with the lattice, these states can have extremely long spin coherences[12,69,136] and lifetimes[137]. There are two major types of nuclear spins: intrinsic nuclear spins within a defect containing impurity atoms, and extrinsic nuclear spins in the atoms surrounding the electronic defect, mainly the non-zero nuclear spin isotopes of the host crystal.

For intrinsic systems, every defect can deterministically have one or more corresponding nuclear spin registers with the proper choice of isotope during defect formation[138–140]. The hyperfine interaction for intrinsic nuclear spins can be large from the contact term (up to GHz[141]). This usually results in a strongly coupled electron-nuclear spin system with nuclear spin-resolved transitions, depending on the electron spin linewidth $\Gamma = 1/\pi T_2^*$. Importantly, the host crystal can be fully isotopically purified while still retaining this intrinsic register[91,92].

However, to extend beyond one nuclear spin for each defect, extrinsic nuclei are necessary. They are also the only nuclear spins available for vacancy-related defects with no corresponding impurity atom. This can likewise produce a few strongly coupled nuclear spins for the first few neighboring sites in the lattice, depending on $\Gamma$.

There are two broad choices for nuclear spin control: direct resonant driving or conditional phase accumulation, depending on the frequency shift $\Delta\omega$ imparted by the hyperfine interaction with the electron spin (see Fig. 3c). With the presence of this control, nuclear spin initialization by single-shot measurement[69] or by swapping polarization with the electron are possible[134]. In strongly-coupled electron-nuclear spin systems, $\Delta\omega$ is large compared to $\Gamma$ and the nuclear spin states are sufficiently resolved for direct magnetic driving of fully-entangling two-qubit gates[133], consisting of electron-nuclear conditional rotations, as shown in Fig. 3d. Spin selectivity here requires significant nuclear Zeeman interaction (high magnetic field), a quadrupole interaction, or hyperfine differences between spin sublevels.

The second way to mediate two-qubit gates is by creating an electron spin superposition and accumulating a nuclear spin-dependent phase[69] (Fig. 3e). The spin state selectivity is similarly limited by the spin's $T_2^*$ and requires a hyperfine interaction that is faster than the dephasing time. Extending the number of registers to include weakly-coupled nuclear spins with small frequency shifts $\Delta\omega$ can be achieved with dynamical decoupling-based control. This extends the $T_2^*$ limit to $T_2$ by cancelling all interactions with the environment but those at a frequency set by the inter-pulse spacing of the decoupling scheme (Fig. 3e). When this frequency matches that of a nuclear spin transition, the electron spin accumulates a nuclear spin-dependent phase depending on the quasi-unique signature from $\Delta\omega$, selective down to single nuclei in the lattice[142,143]. This technique (and variants with interleaved radiofrequency tones[12,144]) allows both controlled and uncontrolled rotations on the nuclear spin[12].

Both schemes result in gate speeds ranging from a few microseconds to a few milliseconds. This has allowed for: the control of and entanglement with more than 10 nuclear spin registers[12,144], proof-of-principle error correction[133,134], QND readout enhancement[135], and resolving single nuclei at a few nanometers distance[145]. Overall, the total experimental sequence length sets the frequency resolution (from the Fourier transform) of the control and is ultimately limited only by the spin's $T_1$, leading to a trade-off between the number of resolvable registers and the total gate time[91]. For weak hyperfine interactions, the number of possible nuclear spins grows greatly as the allowed gate time is increased, defined by the large ratio of $T_1$ to $T_2^*$.

The interplay between the hyperfine interaction strength and the isotopic abundance is critical (Fig. 3f)[91]. A strong hyperfine interaction allows for fast gate times and even longer coherence times as flip-flops between nuclear spins can be suppressed by the hyperfine-induced detuning $\Delta\omega$ (known as a "frozen-core")[146]. Controlling a nuclear spin species that has a different gyromagnetic ratio than the bath spins also avoids this channel of decoherence at high magnetic fields. Isotopic purification increases both the electron and nuclear spin coherence times, yet reduces the availability of nuclear spins that can be used. Conversely, at high nuclear spin concentrations, the ability to resolve nuclear spin transitions is limited by spectral crowding[91,133] that reduces the two-qubit gate fidelities.

Similar to $T_1$ relaxation, protocols that involve optical excitation of a defect will randomize the electron spin state or hyperfine interaction during illumination, and therefore modulates the nuclear spin frequency and reduces its coherence[69,131,147]. Generally, the robustness of the nuclear spin can be described by the decay of the state fidelity $F$:

$$F = \frac{1}{2} + \frac{1}{2^{N+1}}\left(1 + e^{-\frac{1}{2}(\tau\Delta\omega)^2}\right)^N \qquad (5)$$

where $\tau$ is the timescale of the uncontrolled dynamics and N is the number of uncontrolled events[131,147,148]. The fidelity is exponentially improved by reducing $\Delta\omega$. This makes nuclear spins with low hyperfine interactions (~kHz) desirable for nuclear-assisted QND readout performance or quantum communication[131] where many optical cycles or gates on the electron are required. Electrical readout schemes that rely on ionizing defects can similarly dephase nearby nuclear spins through modulation of the hyperfine interaction[149].

In summary, the defect impurity and the atoms in the lattice determine the availability and performance of nuclear spins to manipulate. From an engineering perspective, delta-doping nuclear layers[93], hetero-nuclear crystal hosts[57], and isotopic engineering[91,150] are promising tools.

In order to use quantum registers for solid-state spins, important properties are:

- Intrinsic nuclear spins for quantum sensing
- Many extrinsic nuclear spins for quantum computation and communication
- A careful choice of the isotopic concentration to balance coherence and nuclear registers
- Nuclear spins resilient to electron spin control, that can be read-out in a single-shot, and that have long coherence times.

*Conclusions*

The properties of electron and nuclear spins provide some of the most compelling advantages for spin qubits in solid-state defects. Important considerations relate to the host material, including isotopic purity, defect concentration, and material properties (e.g. piezoelectricity, Debye temperature). The details of the ground (and sometimes excited) state spin Hamiltonian define the possible control and decoherence mechanisms. The incredibly long spin lifetimes and coherence times have shown that these systems are robust quantum memories. Continuous progress toward higher fidelity control of large nuclear spin clusters[12] can lead to small quantum computers or single high-quality logical qubits.

## Optical properties

The optical addressability of many spin defects provides a photonic interface for quantum applications, driving materials research for defects with optimal optical properties. Though electrical (see section Spin-charge interface) and other methods[70] exist to initialize and readout the spin state, an optical interface is generally desired for its practical ease of use and for isolating single defects. An efficient interface requires understanding the major optical parameters of a spin defect, including the emission spectrum, quantum efficiency, and spin-dependent optical contrast.

*Optical emission and excitation*

The emission wavelength is a basic property that influences factors like spectral transmission and required detector technology. For quantum communication, infrared photons are preferred to reduce optical fiber losses as shown in Fig. 4a. This has driven interest in defects that emit in the telecom band including erbium (1536-1550 nm)[48] and vanadium dopants (1280-1390 nm)[60]. Likewise, applications in biosensing are optimized around 1000 nm in the "second biological optical window" dictated by low photon absorption in salt water[151]. For integrated, low-cost applications, silicon detectors operate optimally at 400-1000 nm, while high quantum efficiency can be achieved with superconducting nanowires into the infrared, at a higher cost.

A desirable emission wavelength can be engineered by sum and difference frequency conversion using non-linear optics such as spontaneous parametric down-conversion (SPDC) and spontaneous four-wave mixing (SPWM), often in a periodically poled material[152]. When mixing single photons however, Raman scattering and other sources can produce noise in the output band of interest, depending on the chosen input pump wavelengths[153]. Importantly, these techniques can preserve the phase coherence of the input photons, which is required for entanglement protocols[152,154].

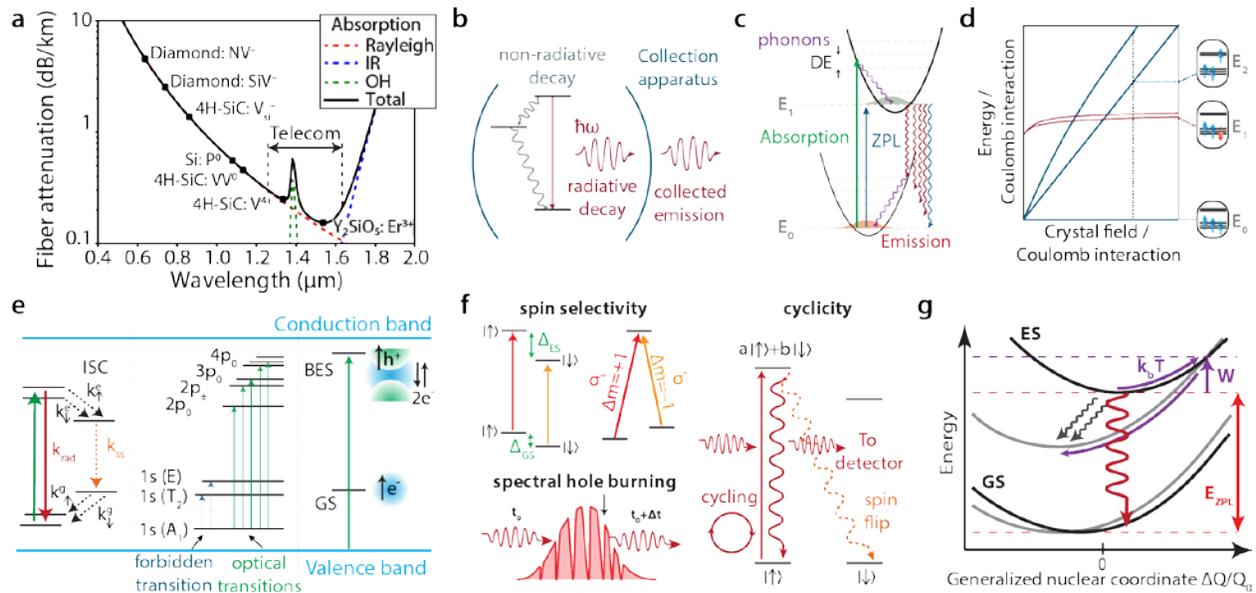

Figure 4 | **Optical properties of spin defects in the solid-state. a|** Attenuation loss as a function of wavelength in an optical fiber. Typical defect emission wavelengths are shown to compare their viability for quantum communication (without frequency conversion). **b|** Radiative and non-radiative emission of defect in a collection system such as free-space optics or photonic cavities. **c|** Phonon-assisted and zero-phonon line (ZPL) absorption and emission from defect orbitals, showing the harmonic vibration potentials in a Franck-Condon diagram. **d|** Tanabe-Sugano diagrams for predicting orbital structures in transition metal ions. Here is shown a $d^3$ orbital with octahedral coordination (e.g. $Cr^{3+}$:MgO), with corresponding energy level and spin assignments for $t_{2g}$ and $e_g$ states on the right. **e|** Common electronic orbital level ordering for defects: from left to right, inter-system crossing (ISC) for spin polarization and readout, a dopant's (e.g. in silicon) many hydrogen-like excited states, ground state (GS) and associated bound-exciton states (BES) of a donor or acceptor. For the sake of simplicity, the label valence and conduction

bands in the manuscript are, respectively, hole and electron ionization thresholds, and normally only coincide for single particle levels. **f|** Spin-dependent optical transitions: (left) spin-resolved optical transitions from frequency detuning or (right) a non-conserving spin transition using polarization selection rules. Spectral hole burning in ensembles for controlled single photon storage. Cyclicity is defined by the spin mixing in the excited state which causes spin flip transitions. **g|** Non-radiative effects from phonon relaxation and its influence on spin-dependent optical processes that can exist in some defects with an ISC (grey arrows). The triplet (black) and singlet (grey) potential energy surfaces are shown with respect to normalized displacements along a harmonic vibrational mode ($\Delta Q/Q_0$). Nonselective transitions (purple) can proceed which characteristic energy $W$.

The photon emission rate is a critical aspect of optically active defects since it defines the experimental signal intensity, the photon count rate of single photon sources, the sensitivity of the defect as a quantum sensor, and the entangling rate for quantum communication. The photon count rate is dictated by the radiative lifetime of the excited state, but is reduced by experimental collection efficiencies, typically < 1% in confocal microscopes (Fig. 4b)[155].

Efficient collection can be addressed via solid immersion lenses and surface metalenses that reduce the effect of total internal reflections, and even collimate the emitted light leaving the sample for efficient free-space collection[155,156]. Alternatively, photonic waveguides can enable direct fiber coupling of the light[8,157,158]. The photon generation rates can be enhanced by using small mode-volume, low-loss cavity structures with a high Purcell factor to reduce the excited state lifetime. Practically, photonic structures can be fabricated directly into the material[8,158–160], deposited on top[157], or made through flip-chip device integration[48].

The emitted photons are spectrally divided into a narrow zero-phonon line (ZPL) and a broad phonon sideband (PSB) (Fig. 4c). The observed emission and absorption spectrum is typically interpreted using a Huang-Rhys model (with predictable factors)[161,162] that provides information about the vibrational structure of the defect's luminescence band[163]. The Debye-Waller factor (DWF) is the key quantity that describes the ratio between the ZPL emission intensity and the overall emission intensity. A low DWF is often due to a strong phonon coupling resulting in decay between the excited state and the higher phonon modes of the ground state[164]. Applications that require photon coherence or interference benefit from a dominant narrow ZPL spectral contribution (high DWF) which contains indistinguishable photons. Nanophotonic cavities provide an engineering pathway for increasing the ZPL emission and the DWF[165,166].

Quantum efficiency (QE) is another key parameter for assessing the performance of an optical emitter. QE is the fraction of excitation events that result in the emission of a photon, and is lowered by non-radiative and ionization rates. The QE can be obtained indirectly by comparing to radiative rates calculated with density function theory (DFT[167]), or by combining master equation modeling and experimental spin-dependent transient decays[168]. Direct experimental measurement of the QE can be achieved by controllably varying the photonic density of states[169].

The radiative and non-radiative decay routes depend on the selection rules such as conservation of total spin $S$, the symmetry of the defect, and the strength of the optical transition dipole moment, spin-orbit coupling and electron-phonon coupling[170]. The symmetry of the defect complexes and their orbitals are generally not known, except for simple cases like the effective mass theory (e.g. donors in silicon)[171]. From the highest symmetry set by the host crystal[26], the defect point group symmetry is lowered for defect complexes with Jahn-Teller distortions, or with applied strain. With significant spin-orbit interactions where $J$ becomes the relevant quantum number, normally "forbidden" transitions become weakly allowed with long optical lifetimes[172]. Other considerations include centrosymmetric defects where transitions are allowed when the starting and ending molecular orbitals have opposite parity, for example[173].

Single particle orbitals and levels for particular defects in a host crystal can be computed by DFT or higher levels of theory (e.g. GW[174]). Molecular orbitals are constructed from these single particle orbitals according to the number of electrons and holes, Jahn-Taller distortions, Hund's and Pauli's rules, and the degeneracy of the orbitals (and labelled according to the overall symmetry and spin multiplicity). The single particle orbital degeneracy directly sets the possible spin multiplicities. $S > 1/2$ systems require degenerate orbitals, for example. The filling and ordering of single particle orbitals at a given charge state can be carried out by DFT and post-DFT calculations (such as quantum embedding theory[167,175]) to construct the defect's molecular orbitals.

Depending on the defect, the optical properties may or may not be affected by the material host. To one extreme, the intra-f-shell transitions in rare earth ions are largely decoupled from their site location and the details of the host lattice, and therefore are transportable across multiple materials. The optical properties of transition metal defects are similarly confined to the d-orbitals of the ion. However, these orbitals are sensitive to the coordination, symmetry, and crystal field strength that the ion experiences. If two hosts share similar properties (e.g. symmetry) then the resulting d-orbital physics is translatable across platforms[172] and can be understood directly through Tanabe-Sugano diagrams (see Fig. 4d)[176,177]. For vacancy-related defects, the details of the resulting dangling bonds and their interaction and relaxation within the lattice define the major optical properties of the defect. Finally, the hydrogenic states of bound excitons are largely determined by the position of their energy levels in relation to the bandgap, not by the details of the internal defect structure, and simplified selection rules from atomic physics can be used[171] (Fig. 4e).

In summary, optical emission properties play a significant role in the practical implementation of defects. A few key guidelines include:

- The need for telecommunication wavelengths for low loss fiber transmission for quantum communication.
- High Debye-Waller factor and high quantum efficiency.
- Optical devices to increase collection efficiency and photon count rates.
- Importance of defect symmetry in determining the spin and optical structures.

*Spin-photon interface*

The spin-dependent optical processes of defects are essential features for spin-based quantum information. Without them, light cannot be used to polarize, control, or readout spins. There are a few ways that the spin of a defect can influence the coupling to light: 1) optical transitions with spectrally or polarization-resolved spin states, or 2) via an intersystem crossing with spin-dependent non-radiative processes.

In the first case, optical polarization (linear or circular) leads to different optical selection rules depending on the spin state[178], and hence to spin-dependent excitation and emission of photons. In the second case, spin selectivity is obtained when two or more spin states in the ground state have different optical transition frequencies to one or more spin states in the excited state (see Fig. 4f)[27]. This frequency resolution requires that the optical linewidth be narrower than the spin-dependent optical shift.

Spin-dependent frequency differences between orbital ground and excited states can arise from different electronic wavefunctions and corresponding spin interactions. This includes difference in the g-factor (e.g. in SiV in diamond[179] and Er-based systems[83]), hyperfine, or zero-field tensors (for $S > 1/2$ such as the NV center in diamond[19] or vacancy centers in SiC[168,180]). A total spin change in the excited state, as with some transition metal systems[181], results in a controllable Λ-system with spin-selective transitions.

For many applications, the spin eigenstates must remain identical between ground and excited states, otherwise the defect would suffer from non-spin-conserving radiative decay from the excited state (Fig. 4f). A low spin-flip probability is desired to create so-called "cycling" transitions, where upon excitation and emission of a photon, the defect is returned to its initial spin state[27]. With this, photons can be continuously extracted from the defect with high correlation to the spin state, critical to spin-photon entanglement for quantum communication protocols[130,178].

Unfortunately, many of the mechanisms that provide spin-selective optical transitions can also contribute to mixing of the spin eigenstates. For example, in some diamond and SiC defects, the axial spin-orbit term $\lambda_z$ creates the frequency splitting needed for spin-photon entanglement, while the transverse spin-orbit mixing $\lambda_\perp$ degrades the spin-selectivity of the transitions and reduces the cyclicity[19]. Additionally, mixing of the spin eigenstates can be reduced/increased by controlled/uncontrolled magnetic fields, electric fields and strains[166,182,183], while Purcell enhancement can increase the cyclicity by changing the balance between different radiative rates[165].

Conversely, non-spin-conserving transitions are required to optically polarize and control the spin ground state. Related to optical pumping in atomic physics, pumping on a transition will eventually result in a spin flip, due to finite cyclicity, and thus in the selective depletion of one spin state[60,84]. This results in a high degree of spin polarization, beyond the thermal Boltzmann distribution, provided the spin $T_1$ is long compared to the spin-flip rate. The optical interface is the key feature that allows operation temperatures that exceed $k_b T$. The cyclicity trade-off between readout and polarization remains a fundamental challenge.

In ensembles, even if the spin-selective structure is inhomogeneously broadened and unresolved, optically pumping the population from a specific spectral band can create spin polarization. These "spectral holes" can be recovered by microwave excitation or by optical means, probing the internal structure within the broadened ensemble[184]. The characteristic time by which the hole recovers is a measure of the spin $T_1^{60,84,181}$. For rare-earth systems, creation of a "comb" of spectral holes allows the ensemble to act as a controllable quantum memory for single photons[185] (Fig. 4f).

Without spin-selective optical excitations, the alternative pathway for optical spin readout and polarization is via spin-dependent non-radiative processes, specifically intersystem crossings (ISC) present for $S > 1/2$. ISC's are non-radiative transitions between orbital levels with different spin multiplicity but the same symmetry, and they are mediated through phonons and the spin-orbit interaction (λ)[186]. Because vibronic mixing and Jahn-Teller distortions in the excited states can mix the different orbitals, resulting in different symmetries, the associated spin projections can couple differently to the ISC through the spin-orbit interaction[187]. Multiple states may be desired in the ISC decay channel to closely match the energies of the ground and excited states for appreciable ISC rates. In general, the ISC rate goes as $\lambda^2 F(E)$ using Franck-Condon theory, where $F$ is the phonon overlap spectral function at the energy spacing $E$ between the levels mediating the ISC[188]. Phonons drive the configurational change of the atoms in the defect for the ISC, whose exact potential energy surfaces (PES) determine the dynamics. For systems with strong electron-phonon coupling, extensions to vibronic states analogous with Herzberg-Teller theory of the optical spectrum[187,189] can also determine the weakly allowed optical transitions within the ISC. The exact mechanisms for the ISC's for various defects are a subject of extensive theoretical and experimental work[167,168,188–190].

The addition of spin-dependent non-radiative rates, as well as intermediate orbital "shelving" states in the ISC, results in spin-dependent emission probability of a photon during the optical cycle (Fig. 4c). This core feature allows for optically detected magnetic resonance (ODMR) of the spin, especially for room temperature sensing applications.

The spin-dependency of non-radiative processes depends on the temperature of the system. For the ISC, the PES of the states for each spin multiplicity can be calculated as a function of the nuclear coordinates of the defect. If an energy barrier ($W$) exists between PES with different multiplicity, then direct transitions between these PES at their crossing point are unlikely[191]. Thermal crossing of the barrier, however, can destroy the spin dependence of the non-radiative ISC rates while also reducing QE[191] (Fig. 4g). The ability to calculate $W$ is thus crucial to design room temperature qubits. Thermal activation through this process modifies the radiative lifetime $\tau$ as a function of temperature $T$ as follows[191]:

$$\tau(T) = \frac{\tau_0}{1 + s \cdot e^{-\frac{W}{k_b T}}} \tag{6}$$

Where $\tau_0$ is the optical lifetime at zero temperature and $s$ is the ratio of non-radiative to radiative rates at the PES crossing point. Spin polarization and measurement contrast using the ISC depends on the relative rates and lifetimes of the system[192], and by knowing all these rates, the optical illumination and readout durations can be optimized for the highest ground state spin polarization (up to 96% [168]), contrast (up to 30% [192]), and number of extracted spin-correlated photons (relating to the signal-to-noise ratio). Finally, it is important to note that the properties yielding a desired ISC will often conflict with an ideal radiative and spin-conserving spin-photon interface. Additionally, the mixing from the presence of an ISC by definition reduces the state purity of the defect's spin.

In sum, the following defect properties are desired:

- A mechanism that maps the qubit spin state to a property (polarization, time, energy) of light constituting spin-photon entanglement.
- Both cycling and non-cycling transitions for single shot readout and efficient spin polarization within the spin relaxation time.
- For the ISC, singlet states should be close in energy to the ground and excited states, high spin selectivity between the ISC rates into or out of the ISC is desired, and a large $W$ for elevated temperature operation

*Optical coherence*

Aside from ground state spin coherence, the coherence of the orbital transitions has important consequences for quantum technologies. Reduced optical coherence can be detrimental to resonant readout protocols[183], spin initialization, optical spin control[126], and signal-to-noise ratio for quantum sensing[193]. Critically, the coherence of an emitted photon directly influences the fidelity and entanglement rate of quantum communication protocols[194,195]. These considerations result in certain applications where optimal optical performance is chosen over better spin properties[8].

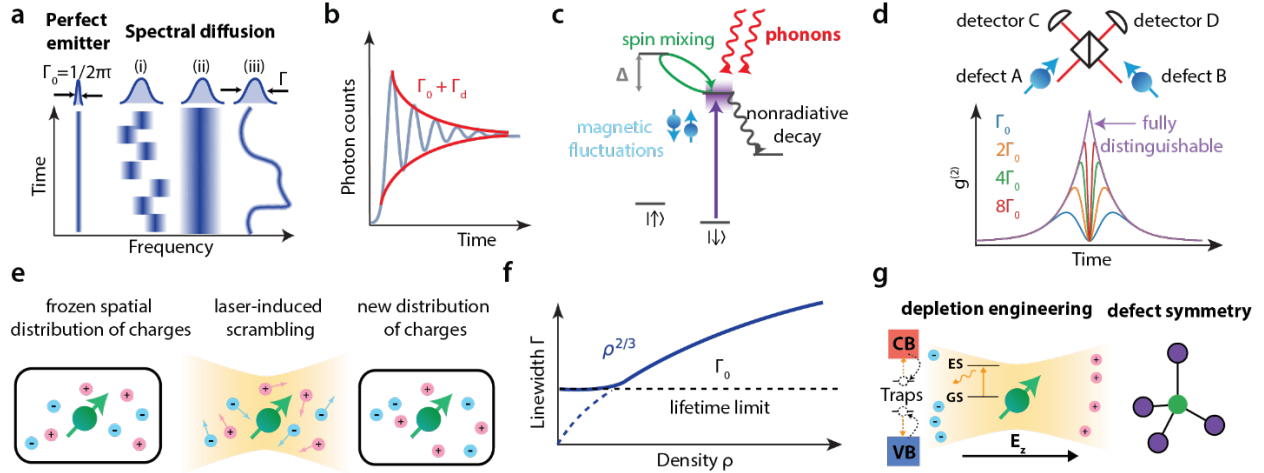

Figure 5 | **Considerations of optical coherence. a|** Various type of spectral diffusion depending on the noise correlation time. **b|** Decay of optical Rabi oscillations through time resolved fluorescence. **c|** possible mechanisms for orbital decoherence and decay. **d|** Hong-Ou-Mandel interference between emitted photons from two defects at a beam splitter. A correlation dip is formed for indistinguishable emitters and is degraded with increased linewidth $\Gamma$. **e|** Mechanism for spectral diffusion through photoionization of charge traps. **f|** Linewidth dependence on charge trap density ρ, with scaling $\rho^{2/3}$ and the lifetime limit $\Gamma_0$. **g|** Possible strategies for mitigating spectral diffusion by: 1) depletion of the fluctuating charges, or 2) engineered defect symmetry for reduced Stark dipole.

For optical transitions, the relevant coherence for single emitters is usually the optical $T_2^*$ or inversely the optical linewidth Γ. An optical emitter with a lifetime τ will have a best-case "lifetime-limited" or transform-limited homogenous linewidth (full width at half maximum) of (Fig. 5a):

$$\Gamma_0 = \frac{1}{2\pi\tau} \qquad (7)$$

This optical linewidth reflects the coherence of the emitted light from the defect, which can also be measured directly through the decay of optical Rabi oscillations[196] (Fig. 5b). Additional non-radiative processes can shorten optical lifetimes and therefore increase Γ (Fig. 5c).

With two perfectly indistinguishable emitters, coherent interference between two emitted photons on a beam splitter will erase the path information of the photons and cause a Hong-Ou-Mandel interference dip[197] (Fig. 5d), enabling heralded entanglement[198]. With imperfect emitters, the reduced phase coherence of the photons reduces the interference visibility and the integration window over which events can be collected[195]. This drastically decreases the entanglement rates and fidelities achievable with defect spins.

A defect's orbital structure can undergo multiple types of broadening, depending on the correlation time of the noise ($\tau_c$) and the noise source. Broadening can manifest as (i) a Gaussian broadened line (short $\tau_c$), as (ii) discrete spectral jumps from experiment to experiment (moderate $\tau_c$), or as (iii) a "spectral wandering" (long $\tau_c$) (Fig. 5a). Phonon-induced optical decoherence occurs through similar processes as discussed in the Spin lifetime section, with direct, Raman and Orbach processes with characteristic temperature scaling[199,200]. In this case, the temperature power law varies between degenerate and non-degenerate orbitals (instead of between Kramers and non-Kramers spin systems) and is dependent on any orbital splitting Δ. Similar to the spin $T_1$, a small spin orbit coupling and large Debye temperature are desired to maintain a lifetime-limited optical line at the desired temperature.

Spectral diffusion (SD) refers to broadening from environmental noise that is usually slow compared to the optical lifetime. For example, thermal drift during the experiment can manifest as slow spectral wandering[199]. Magnetic field fluctuations can broaden and dephase the optical transition, for example in systems with spin-photon interfaces based on a g-factor difference[201] (Fig. 5c). Isotopic purification may not only narrow ensemble linewidths due to mass shifts[202], but also may narrow magnetically sensitive lines[83,203]. Generally, electrical noise tends to be the dominant source of SD and broadening for defect energy levels. If a defect does not have inversion symmetry, it will have a non-zero first order Stark shift dipole that shifts the energy levels due to applied electric fields, usually in different amount between ground and excited levels[180] (see section Optical emission and excitation). Electric field fluctuations are often caused by surface defects[65], through photoionization of nearby impurities[204–206], or by tunneling to a nearby charge reservoir (Fig. 5e). For charge noise from impurities in semiconductors, the total broadening from SD can be understood from both theoretical treatments and Monte-Carlo simulations of charge noise[207], where even high-quality semiconductor crystals with part-per-billion density ($\rho$) of impurities and traps cause significant broadening ($\Gamma \propto \rho^{2/3}$, Fig. 5f). This imposes strict criteria on the purity and quality of the host crystal after growth and defect formation.

Most defect's optical transitions are tunable[206] and dephased by electric fields[204]. Defects with inversion symmetry or exhibiting the same shift in the ground and excited states' levels[180] can display near-transform limited optical lines, even in nanostructures, and can still be tuned with strain fields[59]. Strain tuning should not break the inversion symmetry, otherwise the protection is degraded, and sensitivity to strain must not be so large as to cause additional broadening mechanisms. An important metric is therefore the ratio of optical frequency tuning range to the static inhomogeneous optical linewidth (e.g. caused by different local strain or electric fields). For long distance entanglement where photons need to be indistinguishable[15,208], the achievable tuning range of a defect should be larger than the distribution of optical frequencies. For example, tuning over this range should be possible before reaching dielectric breakdown or mechanical failure.

Besides growth of high purity material, the fluctuating charges can be depleted, turning an electrically noisy environment into a clean one while maintaining Stark sensitivity[204] (Fig. 5g). Fast frequency tuning of the optical lines can allow feedback techniques to stabilize and reduce the linewidth[209]. On the other hand, large Purcell factor enhancements may be beneficial by causing an increased homogenous linewidth that overwhelms the environmental noise without limiting interference visibility.

Systems with high Debye temperature, low spin orbit coupling, of high symmetry, and which exist in clean materials are desirable for optimal optical coherence properties. Other considerations include the need for:

- A tunable spin-photon interface, while limiting optical decoherence and spectral broadening.
- Transform-limited linewidths and low orbital decoherence at a desired temperature.
- Low impurity concentrations, depleted charge environments, or engineered defect symmetries.

*Conclusions*

The optical properties of spin qubit systems offer a natural spin-photon interface to initialize, manipulate and readout the spin state, as well as to mediate entanglement through photons. Important considerations must be made as to the emission wavelength, spectral line shape, and spin contrast when selecting the appropriate defect for a given quantum application. This choice needs to balance transform-limited linewidths, emission brightness, spectral diffusion, and other physical mechanisms that affect the radiative and non-radiative emission rates. Improving these emission properties through optical engineering can mitigate some of these issues, but comes at the complexity of integrating defects into photonic devices.

## Charge properties

The charge properties of spin defects in the solid-state are fundamental to their reliable use, as well as for electrical readout of the spin state. Studying these properties was instrumental to the development of classical condensed matter and semiconductor physics. However, the focus in quantum science is on specific defects and how they can be used, and not on the overall transport properties, as is often the case for classical electronics. Hence a new set of characterization tools and understanding are required.

### *Charge state*

All well-known spin defects correspond to specific charge states (e.g. the negatively charged NV center in diamond), while a different charge state leads to radically different spin and optical interfaces. It is therefore critical that the charge state be 1) initialized on demand and 2) stable during the qubit operation. Condition 1) ensures that a large fraction of a defect ensemble contributes to the signal and a lower fraction contributes as a dephasing spin environment, both being important aspects for sensing applications[210]. For single spins, a known initial charge state is essential for more complex applications such as single shot readout and deterministic entanglement[198]. The condition 2) of stability prevents an additional $T_1$ relaxation channel and, when it occurs in surrounding defects, reduces electrical and magnetic noises that deteriorate the spin and optical coherences[204].

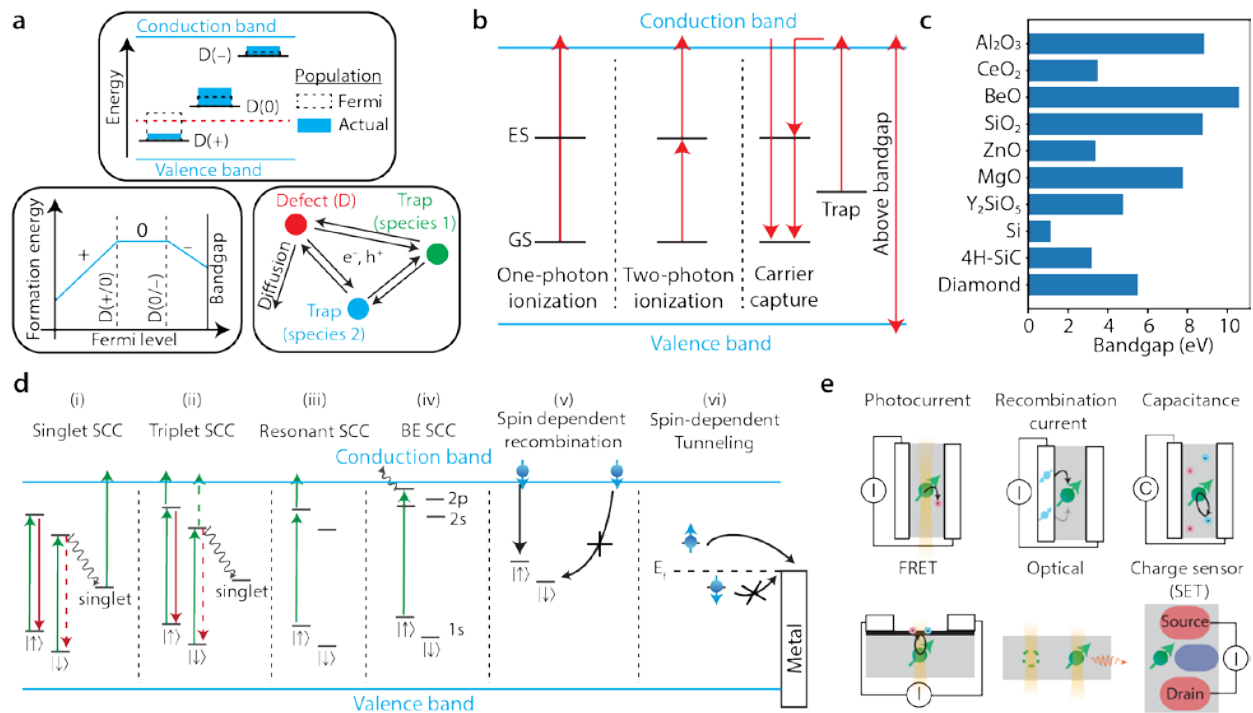

Figure 6 | **Charge state properties of solid-state defects. a**| Top: Population distribution of the charge state in an ensemble of defects (D). At cryogenic temperature under illumination, the population can be far from thermal equilibrium according to the Fermi level (red line). Bottom left: Charge transition energy levels obtained from a formation energy diagram through DFT calculations. Bottom right: Free carrier diffusion, capture and ionization between multiple defect species. **b**| Diagram of typical processes involved in the charge dynamics of defects and free electrons. Equivalent processes are allowed with the valence band for free holes. GS = ground state, ES = excited state. Red arrows correspond to thermal or photo excitation. **c**| Bandgap of relevant host materials for quantum applications[211,212]. **d**| Multiple examples for spin-to-charge conversion (SCC) (i) spin-dependent two photon ionization through spin-dependent shelving (ii) two photon ionization with a spin-dependent lifetime modification (iii) two photon ionization with spin-dependent optical transitions (iv) Bound exciton (BE) SCC with either a two-photon, thermal of Auger ionization process (v) spin-dependent capture of a free carrier to a trap (vi) spin-dependent tunneling. **e**| Electrical readout devices: from top left to bottom right, photocurrent readout after spin-dependent defect ionization, spin-dependent recombination, spin-dependent charge/capacitive sensing after ionization, Förster resonance energy transfer (FRET) between the defect and a secondary system[213], fluorescence readout of the presence or absence of a particular defect charge state, single charge sensor/single electron transistors with spin-dependent tunneling.

Predicting charge stability demands both a theoretical understanding (e.g. from DFT or higher level of theory[214]) and a precise knowledge of the material quality. Electronic structure calculations provide the transition energies between two charge states from formation energy calculations[170], as illustrated in Fig 6a (bottom left). The charge transition level between the charge states 0 and – for example is noted (0/–). The difference in energy between this level and the conduction/valence band is the energy required to remove (– to 0)/add (0 to –) an electron from/to the defect. Using DFT calculations performed for optimized configurations of the defect, optical ionization energies may be inferred from computed charged states by including the Franck-Condon shift, with accuracy depending on the adopted energy functional. The charge transition level for single donors or acceptors are close in energy to the ZPL emission within the binding energy of the exciton. From the electronic wavefunction, the density of states and overall electronic structure of the defect can be calculated to obtain ionization and capture cross-section rates for simulation of charge dynamics[215].

The real complexity in understanding charge stability arises when modeling an actual sample with all relevant impurities. When there is a clear dominant species providing the majority carriers, calculating a Fermi level and transport dynamics may be achievable. However, in samples with large bandgaps and low defect concentration ($10^{13}$-$10^{15}$ cm$^{-3}$), and under illumination at low temperature (carrier freeze-out), charge stability is dictated by the steady state balance between the desired defect and additional electronic traps. Such balance occurs via several kinetic processes (see Fig. 6b) including thermal drift, diffusion, thermal and one or two photon ionization, electron or hole capture, and electron-hole generation from thermal or above-bandgap excitations[216,217].

Practically, the aim is to stabilize the defect to a desired charge configuration by understanding which process occurs under what conditions. There are three main tools for this purpose: electron spin resonance combined with light excitation[218], photoluminescence under two-color illumination[216], and electrical measurements such as deep-level transient spectroscopy (DLTS)[215]. Each tool provides a different signal that is modulated by thermal or optical ionization of the traps, and correlated with temperature, annealing, or sample growth. While exact interpretation can be challenging, combining theory and these experiments can give information such as ionization mechanisms and thresholds, optimal growth conditions, or defect densities[24,215,218].

Some broad considerations can be made solely with respect to the bandgap of the host material, summarized for relevant crystals in Fig. 6c. First, dopants tend to have larger binding energies with lower thermal ionization in wide bandgap semiconductors. Second, the excited state of an optically active defect must be within the bandgap for photoluminescence. For a charge state $q$, the optical excitation energy must be between the $q+1/q$ and $q/q-1$ charge transition energies (with respect to appropriate band edges), and then optimally below a third of the bandgap to avoid two-photon ionization threshold via the excited state, either to the valence or conduction band. Finally, an excitation energy below half of the bandgap is also desirable to avoid mid-gap defects undergoing constant charge cycles via one-photon ionization to and from the valence and conduction bands. These considerations are valid both for the stability of the measured defect and to prevent surrounding impurities from becoming a source of fluctuating noise[204]. A bandgap above 1.6-2.4 eV is therefore optimal for telecom emitters and above 4-6 eV for visible (~600 nm) emitters, for example.

Charge stabilization can be achieved through laser control. If the laser wavelength that excites the optical transition can simultaneously repump to the correct charge state, the defect may be stabilized[206,216]. Multi-color excitations can help create the correct balance of ionization rates for all local traps[216,219]. In addition to optical manipulation, electrical techniques such as depletion engineering or high electric fields can be used to control the amount of local charges and the various ionization and capture rates[204]. Finally, charge control can enable applications such as electrometry[205], super resolution imaging[220], and control of defect formation kinetics[221].

In summary, important considerations are:

- The initialization and stabilization of the charge state for proper qubit function.
- The interplay between the charge state of each defect and the local charge trap environment, under electric fields and photoexcitation.
- A lower bound on the bandgap and knowledge on charge transition levels for charge stability.

*Spin-charge interface*

Readout for defect spins is not limited to ensemble spin resonance (i.e. inductive readout) or purely spin-dependent optical measurements. Spin-to-charge (SCC) conversion allows the spin state of a defect to be mapped to the presence or absence of charges. In the presence of an optical interface, SCC can proceed by the intermediary of spin-dependent optical processes. For defects without an optical interface, SCC can occur via recombination with nearby traps[62], by tunneling to a quantum dot or reservoir of charges[14], or using polarized conduction electrons[222]. These processes are illustrated in Fig. 6d.

We first consider optically assisted SCC, where the large energy scales of optical excitations allow for SCC at elevated temperatures and low magnetic fields. This conversion utilizes spin-dependent photoionization with either spin-selective optical transitions[223] or spin-dependent shelving into metastable states (see section Spin-photon interface)[11,224]. Defects with bound exciton states near a band edge allow for SCC with spin-selective one-photon excitation into these states followed by Auger recombination or thermal excitation[1,225]. One-photon SCC by direct excitation to a band is generally not possible due to the broad transition compared to the spin's energy scales. Instead, two-photon excitation via an excited state is another pathway for SCC where either the first photon to the excited state is spin-selective or the second photon occurs during a spin-dependent ISC decay. Generally, the ionization rate (second photon) should be fast compared to spin reinitialization during optical pumping.

SCC is also of interest for charge-selective optical readout since the switchable charge states have distinct absorption and emission. With low optical power, this readout mechanism can probe the charge state of a defect, originally mapped from its spin state by the SCC, without causing further charge conversion. For defects without a good cycling transition or in cases of low collection efficiency, this method allows for high fidelity measurement and potentially single shot readout[223], or increased signal-to-noise ratio even at room temperature for quantum sensing[11].

SCC without optical assistance requires spin-dependent recombination or tunneling mechanisms. Spin-dependent recombination with local traps, often dangling bonds at interfaces, rely on either thermal polarization of the spin, and thus large magnetic fields and low temperatures, or a relative spin polarization of the defect-trap pair (e.g. pairs are randomly either both spin up or both spin down), even at zero field and high temperature[226]. A related mechanism for both initialization and readout of defect spins is achieved by spin-dependent tunneling to either a reservoir of charges or a quantum dot[14]. This mechanism requires low temperature and high magnetic fields, and is limited by the sharpness of the Fermi-distribution of the electron reservoir[227]. This type of SCC allows for the use of hosts with smaller band gaps through integration with electrical devices.

Independent of the SCC mechanism, electrical readout can occur through a variety of techniques (Fig. 6e). Spin-dependent readout can be directly measured via a photocurrent[228], or from fluctuations in the capacitance of a device due to changes in the trapped charge in a semiconductor[1]. Charge sensors such as single electron transistors can measure single charges[14], so host crystals with this device capability are desired.

Overall, an efficient spin-charge interface requires:

- A high-fidelity spin-dependent ionization process, including optical transitions for high temperature operation and flexibility.
- The ability to deterministically control the charge state of the defect.
- A readout mechanism of defect charge states with single-shot compatibility.
- Engineering capability of the host for integrated electrical devices.

*Conclusions*

Understanding and controlling the charge state of spin defects is critical to their operation in the solid-state. This is achieved by combining knowledge of all impurities in the material with optical and electrical manipulation. Allowed optical transitions for both photoluminescence and ionization strongly influence the charge stability and depend on the position of the defect charge transition levels in the host bandgap. For scalability, electrical readout of the spin state can be achieved using spin-dependent ionization processes.

## Material considerations

Spin defect qubit properties are interwoven with the intrinsic host material properties, including variations in crystallographic, dopant, and nuclear spin imperfections in their local environment. This demands consistently high quality, low strain, and low defect density materials available through the development of mature synthesis and growth processes. Typical growth techniques include chemical vapor deposition and molecular beam epitaxy. More advanced methods like the *in*-situ incorporation of dopants with precise timing during growth and isotopic purification are also useful. The latter is conditional on the availability of nuclear spin-free precursors.

The identification, creation, and localization of defects remains a key challenge to integrate these defects with optical devices, nanostructures, and other spatially dependent applications. The ability to create localized defects must be achieved without introducing significant damage which would negatively affect the defect spin, optical, and charge properties. Additionally, similar defect complexes can often exist in inequivalent lattice sites[50,60,181] and different crystallographic orientations[229], necessitating a detailed understanding of the defect orientation and spin Hamiltonian (e.g. by combining theory, X-ray or ODMR experiments[229]).

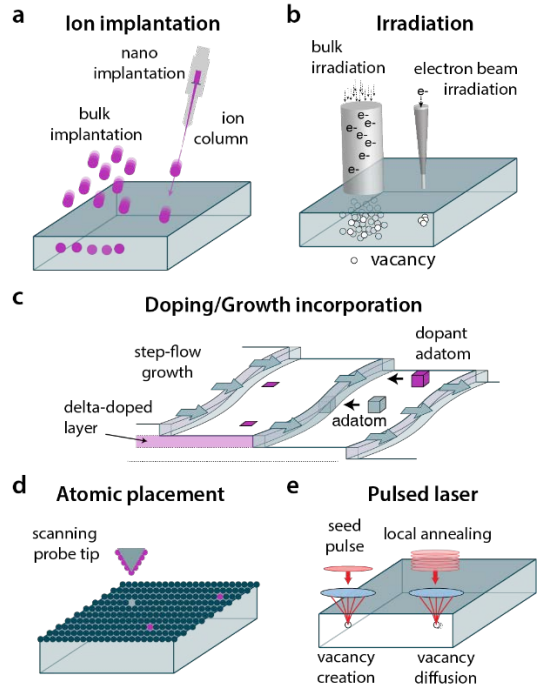

Figure 7 | **Defect host material considerations. a|** Ion implantation can be used to create defects of many different species with a degree of depth localization. Implantation can be done via a bulk process (left) or a nano-implantation process (right) using a modified focused ion beam[230,231]. **b|** Electron irradiation can be used to create vacancies either using bulk irradiation (left) or using an electron beam for focused vacancy creation (right)[232]. While irradiation approaches offer less depth localization, they typically created less damage to the crystal the ion implantation. **c|** Defects can be introduced during growth, which allows for careful control of the defect density, depth, and local nuclear spin environment[139]. Growth methods can incorporate defects while maintaining a high crystal quality in the host material. Figure adapted from Ref.[233]. **d|** Direct placement of atomic defects can be done using a scanning probe tip[234]. **e|** Femtosecond pulse laser systems offers local vacancy creation (single seed pulse at higher laser intensity) and diffusion by annealing (multiple pulses at lower laser intensity).

A common way of creating high quality spin-defects is by starting with low defect density material, and introducing a controlled amount of the desired defects via growth, ion implantation, electron irradiation, or impurity diffusion. Different considerations should be made depending on the defect species:

- Single-atomic defects, including substitutional defects like P:Si, V:SiC and Er:$Y_2SiO_5$, can be introduced via ion implantation or through doping during growth, assuming a suitable precursor growth chemistry is available.
- Atomic-vacancy defects such as NV:diamond and GeV:diamond similarly can be introduced via intentional doping during growth[139,235] or ion implantation[236]. The included vacancy adds a complexity to the defect creation process and can be introduced concurrently with ion implantation or with additional electron irradiation.
- Vacancy complexes such as the $V_{Si}$:SiC and VV:SiC are commonly introduced via electron irradiation (for singles), but can also be created via ion implantation.

Ion implantation is a versatile tool in creating defects (see Fig. 7a) that offers the nearly full periodic spectrum of ions, though at the expense of crystallographic damage to the lattice that can be partially mitigated with subsequent annealing and overgrowth[237]. Spatial (in-plane) localization can be achieved via aperture implantation[236,238] or nanoimplantation[230,231], and the depth can be predicted with Monte Carlo simulations[239]. Electron irradiation uses accelerated relativistic electrons to create vacancies uniformly throughout the sample (see Fig. 7b). These particles primarily create single Frenkel pair displacements in the lattice[240], in contrast to the cascades of defects created by ion implantation. This low-damage method is well suited to single defect creation, with spatial localization achievable using transmission electron microscopes (TEM)[232], compared to bulk accelerators.

For single-atomic and atomic-vacancy complexes[139,241] (see Fig. 7c), a thin epitaxial layer can be grown with controlled doping density from a precursor gas, including isotopic elements, and provides depth localization (within the 'delta-doped' layer) in a high-quality low-strain crystal[242]. For atomic-vacancy defects, the combination of delta-doped growth and localized vacancy creation through aperture implantation[243] or TEM irradiation[232] can give the added benefit of full 3D positioning. Other growth techniques can also control the defect orientation through selective alignment during growth[244,245].

A final critical step is annealing to mobilize dopants and vacancies to create the desired defect. The annealing temperature and time necessary for specific defect formations are dictated by formation energies, migration mechanisms, doping defect densities (both intentional and naturally occurring), diffusion kinetics of impurities and defects, and the structure of the lattice. Likewise, annealing can improve the host's crystal quality and, therefore, the defect properties[246].

Alternative approaches toward localizing defects at the atomic scale have also explored using scanning probe tips[247] (see Fig. 7d) to position dopant defects in silicon[234,248] and 2D materials[249,250]. Such atomic localization becomes critical in device integration and for scaling up spin qubits. Likewise, the use of ultrafast pulsed lasers[251,252], x-ray beams, and focused electron beams[232,253] have been explored to create local vacancies (see Fig. 7b,e). Adding *in-situ* monitoring of the luminescence from optical, x-ray, or electron excitation provides real-time feedback in the defect creation process[253]. Furthermore, using rare isotopes during the creation of atomic-vacancy and substitutional defect complexes can help distinguish ("tag") those defects from naturally occurring ones[139,140]. Finally, localization can also be achieved through integration with nanostructures (nanoparticles[254] and nanopillars[255,256]), embedded particle arrays[257], as well as by charge control and doping within electrical devices[204].

The selection of host materials should consider both the properties of the spin defect, but also the scalability (e.g. wafer availability), ease of fabrication, and unique properties (low acoustic loss, etc.) of the host. In sum, some key parameters are the:

- Availability of high quality, defect-free, wafer-scale and engineerable host material.
- Availability of isotopically purified precursors for growth.
- Existence of a scalable and localized defect creation mechanism (i.e. pulsed laser, irradiation, implantation, growth).

**Outlook**

Starting from initial experiments on spin coherence, every aspect of a defect and its host has now become relevant in the modern quantum context. Where some features may be lacking, the addition of material engineering, e.g. with photonic, phononic or electrical devices, can drastically improve the future viability of the defects.

The interplay between spin, optical, charge, and material properties present a range of trade-offs to consider. The most fundamental balance is between control and coherence of the defect system for any parameter. A good compromise is for low sensitivity to sources of noise but high sensitivity to control and sensed fields. Electron and nuclear spin initialization, coherence, and readout all require a spin-photon or spin-charge interface with opposing requirements for cyclicity (i.e. spin-conservation). Material engineering, such as isotopic dilution or nanofabrication, comes with detrimental consequences such as the loss of nuclear spin registers and increased surface noise or strain.

A variety of novel host materials can be identified according to their nuclear spin concentration, Debye temperature, bandgap, microwave or optical losses, and many more relevant properties as outlined here. Beyond bulk materials, there is a growing interest in 2D materials[17] and molecular qubits[258,259] for bright, localized single photon emitters[249], and new spin and optical tuning mechanisms.

Solid-state spin defects are already being deployed in commercial applications such as in quantum sensing, and the continued progress and understanding of these interconnected properties is vital to fulfilling the full promise of defect-based quantum systems and its future in quantum communication, distributed quantum networks, and other scalable quantum technologies.

## Acknowledgments

We thank Jaewook Lee and Huijin Park for their help in cross-checking the CCE predictions, Hideo Ohno, Tomasz Dietl, Fumihiro Matsukura, and Shunsuke Fukami for fruitful discussion, and Michael Solomon and Grant Smith for reviewing the manuscript. This work was primarily supported by the U.S. Department of Energy, Office of Science, Basic Energy Sciences, Materials Sciences and Engineering Division (G.W., F.J.H, C.P.A., and D.D.A.). H.S. was supported by the National Research Foundation of Korea (NRF) grant

funded by the Korea government (MSIT) (No. 2018R1C1B6008980, No. 2018R1A4A1024157, and No. 2019M3E4A1078666). G.G. was supported by AFOSR FA9550-19-1-0358. S. K. was supported by Marubun Research Promotion Foundation, RIEC through Overseas Training Program for Young Profession and Cooperative Research Projects, MEXT through the Program for Promoting the Enhancement of Research Universities, and JSPS Kakenhi Nos. 19KK0130 and 20H02178. A.G. was supported by the Hungarian NKFIH grants No. KKP129866 of the National Excellence Program of Quantum-coherent materials project, No. 2017-1.2.1-NKP-2017-00001 of the National Quantum Technology Program, No. 127902 of the EU QuantERA Nanospin project, No. 127889 of the EU QuantERA Q_magine project, and by the European Commission of EU H2020 Quantum Technology Flagship project ASTERIQS (Grant No. 820394) as well as the EU H2020 FETOPEN project QuanTELCO (Grant No. 862721).

## Author contributions

All authors contributed to the preparation of the manuscript.

## Competing interests

The authors declare no competing interests.